\documentclass[10pt,onecolumn,conference,letterpaper]{ieeeconf}
\usepackage{amsmath,amssymb,euscript}
\usepackage{latexsym,graphicx,color,amstext,bm,balance}
\usepackage{subfig}

\IEEEoverridecommandlockouts

\graphicspath{{./},{./figs1/.}}

\newtheorem{thm}{Theorem}

\newtheorem{remark}{Remark}

\newcommand{\Var}{\Sigma}

\newcommand{\cp}{\Sigma}
\newcommand{\mW}{{\mathcal W}}
\newcommand{\ud}{{\rm{d}}}

\newcommand{\Expect}{{\mathbb{E}}}

\newcommand{\interior}[1]{%
	{\kern0pt#1}^{\mathrm{o}}%
}
\newcommand{\dbar}{d\hspace*{-0.1em}\bar{}\hspace*{0.2em}}
\newcommand{\mR}{{\mathbb R}}

\newcommand{\ignore}[1]{}

\newcommand{\red}{\color{red}}

\newcommand{\black}{\color{black}}

\definecolor{grey}{rgb}{0.6,0.3,0.3}
\definecolor{lgrey}{rgb}{0.9,.7,0.7}

\newcounter{rmnum}

\newcounter{anum}

\title{\huge Harvesting energy from a periodic heat bath\thanks{Supported in part by the NSF under grants 1807664, 1839441, 1901599, and the AFOSR under FA9550-20-1-0029.}}

\author{Rui Fu$^{*\dagger}$, Olga Movilla Miangolarra$^{*\dagger}$, Amirhossein Taghvaei$^{*\dagger}$, Yongxin Chen$^{\ddagger}$, and Tryphon T.\ Georgiou$^{\dagger}$
	\thanks{$^{\dagger}$Department of Mechanical and Aerospace Engineering, University of California, Irvine, CA; \{rfu2,omovilla,ataghvae,tryphon\}@uci.edu}\\
	\thanks{$^{\ddagger}$School of Aerospace Engineering, Georgia Institute of Technology, Atlanta, GA; {yongchen@gatech.edu}}\\
	\thanks{$^*$Contributed equally and AT directed the completion of the work.}
}

\begin{document}
	
	\markboth{\today}{}
	
	\maketitle

	\begin{abstract}
		The context of the present paper is stochastic thermodynamics--an approach to nonequilibrium thermodynamics rooted within the broader framework of stochastic control. In contrast to the classical paradigm of Carnot engines, we herein propose to consider thermodynamic processes with periodic continuously varying temperature of a heat bath and study questions of maximal power and efficiency for two idealized cases, overdamped (first-order) and underdamped (second-order) stochastic models. We highlight properties of optimal periodic control, derive and numerically validate approximate formulae for the optimal performance (power and efficiency). \\[-.07in]
	\end{abstract}
	
	
	\section{Introduction}
	Harvesting energy is one of the principal characteristics of living organisms. It rarely conforms to the setting of Carnot's cyclic \cite{carnot1986reflexions} contact with alternating heat baths, or the physics of the thermocouple with a stationary thermal gradient. Instead, it is the periodic fluctuations in chemical concentrations in conjunction with the variability of electrochemical potentials that provide the universal source of cellular energy \cite{cilek2009earth}. Thus, energy exchange is often mediated by continuous processes and energy differentials, whereas the Carnot cycle reflects the switching mechanics of an idealized engine.
	
	Yet, for more than 200 years, Carnot's gedanken experiment of quasi-static operation and adiabatic transioning between heat bath of different temperatures, has been the cornerstone of equilibrium thermodynamics {{\cite{adkins1983equilibrium}}}. It has provided us with a wealth of ever expanding insights into the nature of the physical world, from the absolute temperature scale to the concept of entropy, the irreversibility enshrined in the second law, and the time arrow that reigns supreme.
	
	Recent attempts to extend the classical theory of thermodynamics beyond equilibrium processes include the subject of stochastic thermodynamics {{\cite{sekimoto2010stochastic,seifert2012stochastic}}}. This subject has a strong control theoretic flavor and has already provided important new insights. These include the Jarzinski equality \cite{sagawa2010generalized} and the interpretation of dissipation via Hamilton-Jacobi theory for underdamped stochastic dynamical models \cite{chen2019stochastic}.
	
	The topic of this paper follows along similar lines. It puts forth stochastic models for non-equilibrium theremodynamic processes in contact with a heat source having periodically and continuously varying temperature. It then explores the question of how to optimize for energy harvesting, both in terms of power and efficiency. This is a stochastic control problem of a somewhat non-traditional nature; the coupling between controlling potential and heat bath renders the models nonlinear. Expressions in closed form are not possible. Thus, we resort to approximation and numerical verification for limiting cases. Conclusions are drawn as to the nature of optimal operation. Related work treating thermodynamic systems in the linear response regime can be found in~\cite{bauer2016optimal}.
	
	The main contributions of our paper are to be viewed within the context of stochastic control. The motivation and inspiration comes from nonequilibrium thermodynamics processes. That natural processes would somehow self-organize, to match driving potentials and optimize efficiency and power, remains speculatory at present. Specific physical processes need to 
	be modeled, validated, and compared, before any definitive statement is made.
	
	The paper develops as follows. In Section \ref{sec:STm} we present certain basic stochastic models of thermodynamic processes. Sections \ref{sec:OD} and \ref{sec:UD} detail our results for overdamped and underdamped models, respectively, and in Section \ref{sec:CR} we discuss future directions and open questions.
	
	
	\section{Stochastic Thermodynamic models}\label{sec:STm}
	We begin by describing the basic model for a {\em thermodynamic ensemble} used in this work. This consists of a large collection of Brownian particles that interact with a {\em continuous periodic heat bath} in the form of a stochastic excitation and are driven under the influence of an {\em external (time varying) potential}. The dynamics of individual particles are expressed in the form of stochastic differential equations. We consider two models in this paper: the under-damped Langevin equation and over-damped Langevin equation. Control actuation is in the form of a time-varying potential that exerts forcing to individual particles.
	
	\subsection{Under-damped and over-damped Langevin equations}
	The underdamped Langevin equations
	\begin{subequations}\label{eq:underdamped-Langevin}
		\begin{align}
		\ud X_t &= v_t \ud t\label{eq:underdamped-Langevin-x}\\
		m\ud v_t &= \!-\!\nabla_x U (t,X_t)\ud t \!-\!  \gamma v_t \ud t \!+\! \sqrt{2\gamma k_BT(t)}\ud B_t,\label{eq:underdamped-Langevin-p}
		\end{align}
	\end{subequations}
	represent a standard model for molecular systems interacting with a thermal environment. 
	Throughout, $X_t\in \mR^d$ denotes the location of a particle and $v_t\in \mR^d$ denotes its velocity at time $t$, $U(t,x)$ denotes a
	($C^1$ in $t$ and $C^2$ in $x$)
	 time-varying potential for $x\in \mR^d$, $m$ is the mass of the particle, $\gamma$ is the viscosity coefficient, $k_B$ is the Boltzmann constant, $T(t)$ denotes the temperature of the heat bath at time $t$, and $B_t$ denotes a standard $\mR^d$-valued Brownian motion.

	When the inertial effects in the Langevin equation~\eqref{eq:underdamped-Langevin-p} are negligible, specifically, when the temporal resolution $\Delta t \gg \frac{m}{\gamma}$, averaging out the fast variable $v_t$ leads to the {\em overdamped Langevin equation}
	\begin{equation}
	\ud X_t = - \frac{1}{\gamma} \nabla_x U (t,X_t)\ud t + \sqrt{\frac{2k_BT(t)}{\gamma}}\ud B_t.
	\label{eq:overdamped-Langevin}
	\end{equation} 
	Formally, the overdamped Langevin equation is obtained from \eqref{eq:underdamped-Langevin-p} by setting $m = 0$ and replacing $v_t\ud t = \ud X_t$. For a more detailed explanation see~\cite[page 20]{sekimoto2010stochastic}.
	
	In this work we focus on a tractable bilinear model that consists of a quadratic potential and a sinusoidally varying temperature of a heat bath, that is,
	\newcommand{\DT}{{T_1}}
	\begin{align*}
	U(t,x)&=\frac{1}{2}q(t)x^2\\
	T(t)&=T_0+ \DT \cos(\omega t),\;\;\;\;T_0> \DT \geq 0,
	\end{align*}
with $T_0$ and $T_1$ specified.
	
	The state of the thermodynamic ensemble is identified with the  the joint probability density of $X_t, v_t$, denoted by $\rho(t,x,v)$, for the under-damped case, and probability density of $X_t$, denoted by $\rho(t,x)$, for the over-damped case. The corresponding
	Fokker-Planck equations are
	\begin{subequations}
		\begin{align}
		\frac{\partial { \rho}}{\partial{ t}}
		=&-\nabla\cdot \left\{\begin{bmatrix} v \\  -\frac{1}{m}(\gamma v+\nabla_x U) \end{bmatrix} \rho\right\}
		+\frac{\gamma k_BT(t)}{m^2} \Delta_{v} \rho,
		\end{align}	
		and
		\begin{equation}
		\label{eq: Fokker-Planck overdamped}
		\frac{ \partial{\rho}}{\partial{t}}=\frac{1}{\gamma}\nabla_x\cdot \left[ (\nabla_x U +k_BT\nabla_x \log \rho)\rho\,\right],
		\end{equation}
	\end{subequations}
	respectively.
	
	\subsection{Internal energy, heat and work}
	\newcommand{\supu}{{\rm \bf u}}
	\newcommand{\supo}{{\rm \bf o}}
	The internal energy of a single particle, governed by the under-damped Langevin equation~\eqref{eq:underdamped-Langevin},  is the summation of the kinetic energy and the potential energy,
	\begin{subequations}
		\begin{equation}
		E^\supu_t = \frac{1}{2}mv_t^2 + U(t,X_t).
		\end{equation}
		For the over-damped model~\eqref{eq:overdamped-Langevin}, the kinetic energy is negligible and hence ignored, and the internal energy is
		\begin{equation}
		E_t^\supo = U(t,X_t).
		\end{equation}
	\end{subequations}
	The superscript in the notation suggests the case.
	
	Evolution of the thermodynamic ensemble under the influence of the time-varying thermal environment and the time-varying potential $U(t,x)$, leads to an exchange of work and heat. These can be defined at the level of a single particle as explained below.

	The energy exchange between an individual particle and the external potential represents {\em work}. Specifically, the work transferred to the particle by an infinitesimal change in the actuating potential is
	\begin{align}
	\dbar W &= \frac{\partial U}{\partial t}(t,X_t) \ud t. \label{eq:dW}
	\end{align}
	Here we use $\dbar$ to emphasize that $\dbar W$ is not a perfect differential, in that
	  $\int \dbar W$ depends on the path and not just on the end-point conditions. The same definition for work holds for both under-damped and over-damped models. 
	
	The energy exchange between an individual particle and the thermal environment represents {\em heat}. The heat exchange is defined in such a way so that the {\em first law of thermodynamics}, $
	\ud E_t=\dbar Q + \dbar W 
	$ holds. Because the internal energy in the over-damped model does not involve the kinetic energy, the heat is different for the two models. It is 
	\begin{subequations}
		\begin{align*}
		\dbar Q^\supu&= \nabla_xU(t,X_t) \circ \ud X_t=- \gamma \|v\|^2\ud t + \frac{k_BT(t)\gamma}{m}\ud t+ \sqrt{2k_BT(t)\gamma}v\ud B_t,
		\end{align*}
		for the under-damped model, and
		\begin{align*}
		\dbar Q^\supo=& \nabla_xU(t,X_t) \circ \ud X_t\\
		=&- \frac{1}{\gamma} \|\nabla_x U (t,X_t)\|^2\ud t +  \Delta_x U(t,X_t) \frac{k_BT(t)}{\gamma}\ud t+\nabla_x U (t,X_t) \sqrt{\frac{2k_BT(t)}{\gamma}}\ud B_t,
		\end{align*}
	\end{subequations}
	for the over-damped model. In each case, the first entry on the right is the expression in the Stratonovich form, whereas the second entry brings in the correction term for changing into the It\^o form. For details see~\cite[section 4.1]{sekimoto2010stochastic}.  
	
	Accordingly, for a thermodynamic ensemble at a state $\rho(t,x,v)$ or $\rho(t,x)$, the work and heat differentials are obtained by averaging over the ensemble,
	\begin{subequations}
		\begin{align} \label{eq:dbbW2}
		\dbar \mathcal W^\supu &= \left[ \int \int \frac{\partial U}{\partial t}(t,x)\rho(t,x,v)\,\ud x\ud v\right]\ud t,\\
		\label{eq:dbbQ2}
		\dbar \mathcal Q^\supu &=\left[\!  \int\int\!\left(\!-\gamma \|v\|^2\! +\! \frac{k_BT(t)\gamma}{m}\!\right)\rho(t,x,v)\,\ud x\ud v \!\right]\!\ud t,
		\end{align}
		and
		\begin{align}
		\label{eq:dbbW}
		\dbar \mathcal W^\supo &= \left[ \int \frac{\partial U}{\partial t}(t,x)\rho(t,x)\,\ud x\right]\ud t,\\
		\label{eq:dbbQ}
		\dbar \mathcal Q^\supo &=\left[\! \int\!\left(\!-\! \frac{1}{\gamma} \|\nabla_x U\|^2 \!+\! \Delta_x U \frac{k_BT(t)}{\gamma}\!\right)\!\rho(t,x)\,\ud x \!\right] \!\ud t,
		\end{align}
	\end{subequations}
	for the under-damped and over-damped models, respectively. The expressions satisfy the first law of thermodynamics for the ensemble
	$
	\ud \mathcal E(\rho,U) = \dbar \mathcal Q + \dbar \mathcal W,
	$
	where the internal energy is given by
	\begin{subequations}
		\begin{align}\label{eq:dbbE2}
		\mathcal E^\supu(\rho,U) &= \int \int \left(\frac{1}{2} m\|v\|^2
		+U(t,x)\right)\rho(t,x,v)\,\ud x\ud v,
		\end{align}
		and 
		\begin{align}\label{eq:dbbE1}
		\mathcal E^\supo	(\rho,U) &= 
		\int_{\mR^d} U(t,x)\rho(t,x)\,\ud x,
		\end{align}
	\end{subequations}
	for the two models, respectively. 
	
	\subsection{Work and heat exchange with periodic temperature}
	We now consider a cyclic process of period $\frac{2\pi}{\omega}$ in which energy is extracted and heat exchanged from a heat bath with varying temperature. 
	{{Without loss of generality, we restrict to the scalar case, i.e., $d=1$.}} 
	Under the  quadratic potential $U(t,x) = \frac{1}{2}q(t)x^2$, 
	the work extracted over a cycle, $\mW=-\int_0^{\frac{2\pi}{\omega}} \dbar \mathcal W$, can be expressed as (using \eqref{eq:dbbW} or \eqref{eq:dbbW2})
	\begin{align}\label{eq:workcycle}
	\mW&=-\mathbb{E}\left\{\int_0^{\frac{2\pi}{\omega}} \dbar W \right\}
	=-\frac{1}{2}\int_{0}^{\frac{2\pi}{\omega}} \dot{q}(t)
	\mathbb{E}\{X_t^2\}
	\ud t.
	\end{align}

	In contrast to the classical Carnot cycle, where the engine switches contact adiabatically between two heat baths of differing temperatures, such a delineation of phases in the cycle is no longer rigid. It is now the sign of the heat flux differential $\dbar \mathcal Q$ that determines the phase of the cycle when heat flows in, or out of the ensemble, respectively. Thus, accordingly,
	\begin{align}\label{eq:newheat}
	\mathcal{Q}^{}_h&=\int_0^{\frac{2\pi}{\omega}}(\dbar \mathcal Q)_+,\quad 
	\mathcal{Q}^{}_c= - \int_0^{\frac{2\pi}{\omega}}(-\dbar \mathcal Q)_+,
	\end{align}
	specify the heat flowing in and out of the ensemble, in the respective portion of the cycle. Here, $(x)_+=\max\{0,x\}$ for all $x\in \mR$.
We note that alternative definitions of heat have been used, see, e.g., \cite{bauer2016optimal}.

	Our goal is to determine the average power
	$${\rm power}=\frac{\omega}{2\pi}\mathcal{W},$$
	that can be made available by suitable choice of  the linear control gain $q(t)$. 
	Moreover, we are interested in assessing the efficiency $\eta$ of the heat engine while operating at maximum power, namely
	$$\eta=\frac{\mathcal{Q}^{}_h-\mathcal{Q}^{}_c}{\mathcal{Q}^{}_h}=\frac{\mathcal{W}}{\mathcal{Q}^{}_h}.$$
	Those two problems lead us to consider different versions of the control problem to maximize the performance index
	\begin{equation}\label{eq:problem}
	\max_q\{-\frac{\omega}{4\pi} \int_0^{\frac{2\pi}{\omega}} \dot q(t)\mathbb{E}\{X_t^2\}\ud t 
	\},
	\end{equation}
	over choice of the control gain $q(t)$. 
	{{This problem, in the context of a Carnot-like cycle, has already been tackled both within the overdamped \cite{overdampedseifert,fu2020maximal} and the underdamped \cite{optimization2017masters}, \cite{underdampedlutz} framework.}}

	\section{Performance for Overdamped Dynamics}\label{sec:OD}
	\subsection{Analysis for maximal power}
	Consider the over-damped Langevin dynamics~\eqref{eq:overdamped-Langevin}. Without loss of generality, assume the mean of $X_t$ is zero.  Let $\cp(t)\triangleq \Expect[X_t^2]$ denote the variance of $X_t$. The evolution of the variance  is governed by the differential Lyapunov equation
	\begin{align}\label{Lypoverdpd}
	\dot \cp(t)=-\frac{2}{\gamma}q(t)\cp(t)+\frac{2}{\gamma}k_BT(t).
	\end{align}
	We consider sinusoidal temperature fluctuation
	\begin{equation}\label{eq:Toft}
	T(t) = T_0 + \epsilon \DT  \cos(\omega t)
	\end{equation}
	about the mean value $T_0$, 
	and study the potential for drawing power
	by applying periodic control 
	\begin{align}\label{eq:q}
	q(t)=q_0+ \epsilon u(t),
	\end{align}
	where  $\epsilon u(t)$ is the deviation of the control input from the nominal value $q_0$.   
	We study the limit case when perturbations, and thus $\epsilon$, are small.
	
	We seek to determine a control input that maximizes power, that is,
	\begin{align}\label{eq:optpower}
	\max_{u} \quad &- \frac{\omega}{4\pi}\int_{0}^{\frac{2\pi}{\omega}} \epsilon\dot{u}(t)\cp(t)\ud t,
	\nonumber\\\text{subject to} \quad & \dot{\cp}(t)=-\frac{2}{\gamma}(q_0+\epsilon u(t))\cp(t)+\frac{2}{\gamma}k_BT(t),
	\nonumber\\
	&u(0)=u(\frac{2\pi}{\omega}).
	\end{align}
	To this end, we carry out a perturbation analysis about $\epsilon=0$. The variance $\Sigma(t)$ is expressed as
	\begin{equation}
	\cp(t) = \sum_{k=0}^\infty \epsilon^k \cp^{(k)}(t) 
	\end{equation}
	where $\cp^{(k)}(t)$ solves the Lyapunov equation~\eqref{Lypoverdpd} for $\epsilon^k$ order. In particular, the leading two terms satisfy
	\begin{align*}
	\dot{\cp}^{(0)}(t) &= - \frac{2q_0}{\gamma} \cp^{(0)}(t) + \frac{2k_BT_0}{\gamma}\\
	\dot{\cp}^{(1)}(t) &= - \frac{2q_0}{\gamma} \cp^{(1)}(t)  - \frac{2u(t)}{\gamma} \cp^{(0)}(t) + \frac{2k_B\DT }{\gamma} \cos(\omega t).
	\end{align*}
	We truncate all but the first two terms in the objective function of the optimal control problem, and consider
	the problem to optimize
	\begin{align}\label{eq:optpower2}
	\max_u\;\;  - \frac{\omega}{4\pi}\int_{0}^{\frac{2\pi}{\omega}}  \left(\epsilon\dot{u}(t)\cp^{(0)}(t) + \epsilon^2 \dot{u}(t)\cp^{(1)}(t)\right)\ud t.
	\end{align}
	The solution of the optimal control problem~\eqref{eq:optpower2} can now be expressed as follows. 
	
	
	
	\begin{thm}\label{thm:thm1}
		Consider the optimal control problem~\eqref{eq:optpower2}. The optimal control law is 
		\begin{align}
		\label{eq:opt-u-over-damped}
		u^{*}(t)&= q_1^* \cos(\omega t  - \phi^*),
		\end{align}
		where 
		\begin{subequations}\label{eq:q-phi-od}
			\begin{align}
			q_1^* &= \frac{q_0g\DT }{2\gamma \omega T_0} ,\quad 
			\phi^* = \angle (\gamma \omega - i2q_0),
			\end{align}
		\end{subequations}
		with $g = \sqrt{\gamma^2\omega^2+4q_0^2}$,
		giving power output
		\begin{align*}
		\rm{power}&=\epsilon^2 \frac{k_Bq_0\DT ^2}{8\gamma T_0} +O(\epsilon^3).
		\end{align*}
		
		%

	\end{thm}
	\medskip

	For assessing efficiency in next section, we provide here the expression for the variance $\Sigma$ up to first order in $\epsilon$:
	\begin{align}\label{eq:var-overdamped}
	\Sigma(t) = &\frac{k_BT_0}{q_0} + \epsilon \frac{k_B\DT }{\gamma \omega} \sin(\omega t) + O(\epsilon^2),
	\end{align}

	\subsection{Efficiency at maximal power output}
	For the over-damped Langevin model~\eqref{eq:overdamped-Langevin}, the heat exchange rate~\eqref{eq:dbbQ} simplifies to
	\begin{equation}
	{\dbar \mathcal Q} = \frac{1}{2}q(t) \dot{\cp}(t) \ud t.
	\end{equation} 
	Using the control input~\eqref{eq:opt-u-over-damped}, and the variance~\eqref{eq:var-overdamped} the heat exchange rate is 
	\begin{align}
	{\dbar \mathcal Q} =	\frac{1}{2}\left(q_0+\epsilon u(t)\right)\dot{\cp}(t)\ud t
=\frac{k_BT_0q_0}{2\gamma}
	\left(\!\epsilon \frac{\DT }{T_0}\!+\!\frac{\epsilon^2 g\DT ^2}{2\gamma \omega T_0^2}\cos(\omega t\!-\!\phi^*)
	\!\right)\cos(\omega t) \ud t \!+\!O(\epsilon^3).
	\end{align} 
	For $\epsilon$ small, the rate is positive over an interval $t\in[
	t_2-\frac{2\pi}{\omega}, t_1 
	]$ and negative
	over $t\in[t_1 ,t_2 ]$, for the values 
	\begin{equation}
	\begin{aligned}
	\label{eq:t}
	t_1 &=\frac{\pi}{2\omega} +O(\epsilon), 
	\quad
	t_2 = \frac{3\pi}{2\omega} + O(\epsilon).
	\end{aligned}
	\end{equation}
	During these time intervals, the heat exchanged is
	\begin{align*}
	\mathcal Q_h&= \epsilon
	\frac{k_B\DT q_0}{\gamma \omega}\left(1+\epsilon\frac{\pi \DT }{8T_0}\right)
	+O(\epsilon^3),
	\\
	\mathcal Q_c&= \epsilon\frac{k_B\DT q_0}{\gamma \omega}\left(1-\epsilon\frac{\pi \DT }{8T_0}\right)
	+O(\epsilon^3),
	\end{align*}
	respectively. Hence, the efficiency at maximum power is 
	\begin{align}\label{eq:eta-od}
	\eta &=\frac{\mathcal Q_h-\mathcal Q_c}{\mathcal Q_h}= \epsilon \frac{\pi}{4}\frac{\DT}{T_0} +O(\epsilon^2).
	\end{align} 
	
	\begin{remark}
		We note that Carnot efficiency, for quastistatic operation between two heat baths of temperatures $T_h$ and $T_c$, for hot and cold respectively, is $\eta_C=\frac{T_h-T_c}{T_h}$. Letting $T_h=T_0+\epsilon \DT$ and $T_c=T_0-\epsilon \DT$, and evaluating the expression for $\eta_C$, gives 
		\[
		\frac{2\epsilon T_1}{T_0+\epsilon \DT}=2\epsilon \frac{T_1}{T_0} + O(\epsilon^2).
		\]
		The extra factor of $\pi/8$ in \eqref{eq:eta-od} is intriguing, and appears to relate to the sinusoidal shape of the temperature profile that generates an elliptic T-S diagram.
	\end{remark}
	
	
	\black

	\black

	\section{Performance for Underdamped Dynamics}\label{sec:UD}
	\subsection{Analysis for maximal power}
	We now consider the under-damped Langevin model~\eqref{eq:underdamped-Langevin}, subject to heat bath with continuously periodic temperature of the form 
	\begin{align*}
	T(t)=T_0+\epsilon T_1\cos(\omega t),
	\end{align*}
	as before, with $\epsilon$ a small parameter and control gain
	\begin{equation}
	q(t)=q_0+ \epsilon u(t).
	\end{equation} 
	with control input $u(t)$ about the nominal gain $q_0$. As always, the objective is to extract maximum power via a suitable choice of input.  
	We follow the same approach as in the overdamped case and compare our results with the results obtained in Section~\ref{sec:OD} .
	
	Now $\Var(t)$ denotes the $2\times 2$ covariance matrix for the stochastic vectorial process   $(X_t,v_t)^\prime$ that includes position and velocity. It obeys the Lyapunov equation 
	\begin{equation}\label{lyap}
	\dot{\Var}(t)=A(t)\Var(t)+\Var(t) A(t)'+D(t)D(t)',
	\end{equation}
	for  
	\[
	A(t)=\begin{bmatrix}
	0&1 \\
	-\frac{q(t)}{m}&-\frac{\gamma }{m}
	\end{bmatrix},\;
	D(t)=\begin{bmatrix}
	0\\
	\sigma(t)
	\end{bmatrix},
	\]
	and {$\sigma(t)=\sqrt{2\gamma k_BT(t)/m^2}$.}
	Our problem to maximize power for small perturbations $u(t)$ becomes: 
	\begin{align}\label{eq:opt-cont-under}\nonumber
	\max_{u} &\quad -\frac{\omega}{4\pi}\int^{\frac{2\pi}{\omega}}_0 \epsilon \dot{u}(t) \Var_{11}(t)\ud t\\ \nonumber
	\mbox{subject to}&\ \ \dot{\Var}=A\Var+\Var A'+DD'\nonumber\\
	&\ \ u(0)=u(\frac{2\pi}{\omega}).
	\end{align}
	
	We carry out analysis for small values of $\epsilon$. To this end, we consider a series expansion for the covariance,
	\begin{align*}
	\Var(t)&= \Var^{(0)}(t)+\epsilon \Var^{(1)}(t) + O(\epsilon^2),
	\end{align*}
	where $\Var^{(0)}(t)$ and $\Var^{(1)}(t) $ satisfy
	\begin{align*}
	\dot{\Var}^{(0)}(t)&=A_0\Var^{(0)}(t)+\Var^{(0)}(t) A_0'+D_0D_0',\\
	\dot{\Var}^{(1)}(t)&=A_0\Var^{(1)}(t)+\Var^{(1)}(t) A_0'
	+U(t)\Var^{(1)}(t)+\Var^{(1)}(t) U(t)'+D_1(t)D_1'(t),
	\end{align*}
	where
	\begin{align*}
	&A_0=\begin{bmatrix}
	0&1 \\
	-\frac{q_0}{m}&-\frac{\gamma }{m}
	\end{bmatrix},\; 
	D_0=\begin{bmatrix}
	0\\
	\sqrt{\frac{2\gamma k_B T_0}{m^2}}
	\end{bmatrix},\\
	&U(t)=\begin{bmatrix}
	0&0 \\
	-\frac{u(t)}{m}&0
	\end{bmatrix},\;
	D_1(t)=\begin{bmatrix}
	0\\
	\sqrt{\frac{2\gamma k_B T_1}{m^2}\cos(\omega t)}
	\end{bmatrix}.
	\end{align*} 
	Inserting the expansion terms into the objective function~\eqref{eq:opt-cont-under}, and retaining the first two leading terms yields
	\begin{align}\label{eq:opt-cont-under-1}
	\max_{u} &\;- \frac{\omega}{4\pi}\int^{\frac{2\pi}{\omega}}_0 \left[\epsilon \dot{u}(t) \Var^{(0)}_{11}(t)+ \epsilon^2 \dot{u}(t) \Var^{(1)}_{11}(t)\right] \ud t.
	\end{align}
	A 
	variational approach to this maximization problem results in the solution summarized below.
	
	\begin{thm}
		Consider the  optimal control problem~\eqref{eq:opt-cont-under-1} for the underdamped Langevin model. Then, the optimal control law is 
		\begin{equation}
		u^*(t)=q_1^*\cos(\omega t-\phi^*),
		\end{equation}
		where 
		\begin{subequations}\label{eq:q-phi-ud}
			\begin{align}
			q_1^*&=\frac{ q_0 \gamma T_1\sqrt{\alpha^2+\beta^2}}{T_0 (2\gamma \beta-\alpha \omega m) },		\quad
			\phi^*=\angle(\beta-i\alpha),
			\end{align}
		\end{subequations} with 
		$\alpha=4q_0\gamma-3\omega^2\gamma m$, $\beta=\omega(2\gamma^2+4q_0 m-\omega^2 m^2)$, giving power output 
		\begin{equation}\label{power}\mbox{power}=\epsilon^2\frac{ q_0 \omega k_B\gamma^2T_1^2}{2T_0 (2\gamma \beta-\alpha \omega m) }+O(\epsilon^3).
		\end{equation}
	\end{thm} 
	\medskip
	
	For the purpose of evaluating efficiency in the next section, we note that the time variation of the entries of the covariance matrix for the optimal choice of control is as follows
	\begin{subequations}
		\begin{align}
		\Sigma_{11}(t) &= \frac{k_BT_0}{q_0} +\epsilon\frac{2\gamma k_BT_1}{2\gamma \beta -\alpha \omega  m }\left[2\gamma \sin(\omega t) - m \omega  \cos(\omega t)\right]+O(\epsilon^2),\\
		\Sigma_{12}(t) &=\epsilon\frac{\omega \gamma k_BT_1}{2\gamma \beta - m \omega \alpha}\left[2\gamma \cos (\omega t) + m \omega \sin(\omega t)\right]+O(\epsilon^2),\\ 
		\label{eq:Sigma22}
		\Sigma_{22}(t) &= \frac{k_BT_0}{m} - \epsilon\frac{2\gamma k_B T_1r_2}{r}\sin(\omega t +\phi^* + \theta ) +\epsilon\frac{2\gamma k_B T_1q_0\omega}{2\gamma \beta -m\omega \alpha }\cos(\omega t - 2\phi^*)+O(\epsilon^2),
		\end{align} 
	\end{subequations}
	where $r=\sqrt{\alpha^2+\beta^2}$, $r_2=\sqrt{(\frac{2q_0}{m}-\omega^2)^2+(\frac{\omega \gamma}{m})^2}$, and $\theta = \angle \left( \frac{2q_0}{m}-\omega^2+i\frac{\omega \gamma}{m}\right)$. 

	\medskip
	
	\begin{remark}
		In the limit as $m\to 0$,  we have $\alpha \to 4q_0\gamma$, $\beta \to 2\gamma^2 \omega$, and we recover the optimal control law and the power output that were derived in over-damped limit in Section~\ref{sec:OD}. 
	\end{remark}
	
	\subsection{Efficiency at maximal power output}
	For the underdamped model that we consider here, the heat exchange rate is given by:
	\begin{align*}
	{\dbar \mathcal Q} = \gamma \left(\frac{k_B}{m} T(t)- \Var_{22}(t) \right)\ud t .
	\end{align*}
	Using the expressions for $\Sigma_{22}$ in~\eqref{eq:Sigma22}, the rate becomes 
	\begin{align*}
	{ \dbar \mathcal Q} =\epsilon\frac{\gamma k_B T_1}{m}\bigg( &\cos(\omega t) - \kappa_1{\sin(\omega t-\phi^* + \theta)}
	{-\kappa_2\cos(\omega t -2\phi^*)}\bigg)\ud t + O(\epsilon^2),
	\end{align*}
	where $\kappa_1 = \frac{2m\gamma r_2}{r}$ and $\kappa_2=\frac{2m\gamma q_0\omega }{2\gamma \beta  - m \omega  \alpha}$.  The expression can be simplified to
	\begin{align*}
	{\dbar\mathcal Q} =\epsilon\frac{\gamma k_B T_1}{m} \kappa \cos(\omega t + {{\psi}})\ud t + O(\epsilon^2),
	\end{align*}
	where 
	\begin{align*}
	\kappa^2 = &(1-\kappa_1{\sin(\theta-\phi^*) }
	{- \kappa_2\cos(2\phi^*)})^2 + ({-\kappa_1\cos(\theta-\phi^*) - \kappa_2\sin(2\phi^*)})^2.
	\end{align*}
	The precise value of the angle ${{\psi}}$ is irrelevant for the computation of the total heat exchange (hence not included). Integrating over the time interval where the heat exchange rate is positive, yields
	\begin{align}
	\mathcal Q_h = \int_0^{\frac{2\pi}{\omega}} (\ud Q)_+ = \epsilon \frac{2 \kappa \gamma k_BT_1}{m\omega}  + O(\epsilon^2).
	\end{align}  
	From this we readily determine the efficiency as being
	\begin{align}\label{eq:eta-ud}
	\eta =\frac{\frac{2\pi}{\omega } {\rm power}}{\mathcal Q_h} = \epsilon{ \frac{ \pi m  \omega q_0 \gamma T_1}{2\kappa T_0 (2\gamma \beta-\alpha \omega m) } }+ O(\epsilon^2).
	\end{align}

	\begin{remark}
		Once again, as $m\to 0$, we recover the results in Section \ref{sec:OD}. Specifically, as $m\to 0$,  $\kappa_1 \to 1$, $\kappa/m\to \frac{q_0}{2\gamma^2}$. Hence, the heat exchange $\mathcal Q_h \to \epsilon \frac{k_BT_1q_0}{\gamma \omega}$ and the efficiency $\eta\to\epsilon \frac{\pi}{4}\frac{T_1}{T_0}$.
	\end{remark}

	\section{Numerical validation}\label{sec:numerics}
	In this section we provide numerical validation and insight into the effect of higher order terms in the expansions. Specifically, we  
	consider a sinusoidal control input and use Fourier representations to numerically solve the Lyapunov equation and obtain expressions for the power. Our interest mainly focuses on how maximal power depends on the amplitude and phase of the control, and on how efficiency at maximal power depends on the amplitude of the temperature fluctuations of the heat bath.
	
	Starting from the choice of control
	\[
	q(t)=q_0+q_1\cos(\omega t-\phi)=a_{-1}e^{-i\omega t}+a_0+a_1e^{i\omega t},
	\]
	where $a_0=q_0$ and $a_1=\bar a_{-1}= \frac12 q_1e^{-i\phi}$, the power drawn can be expressed as
	{{
			\begin{align}
			\mbox{power}=-\frac{i\omega}{2}\left(a_1c_{-1}-a_{-1}c_1\right)
			\end{align}
			and
			\begin{align}
			\mbox{power}=-\frac{i\omega}{2}(a_1 \Var_{11,-1}-a_{-1}\Var_{11,1}),
			\end{align}
			for overdamped and underdamped dynamics respectively,
			where $c_n$ is the $n$th Fourier coefficient of covariance $\cp$ for overdamped dynamics}}, and $\Var_{11,n}$ is the $n$th Fourier coefficient of 
	$\Var_{11}$ for underdamped one. The Fourier coefficients can be obtained by expressing the Lyapunov equation in the Fourier domain, to obtain a set of linear coupled equations for the various terms. We truncate and keep the first $100$ modes, and solve the resulting finite-dimensional problem for a range of values for $q_1$ and $\phi$.
	
\begin{figure*}[t]
		\centering
		\begin{tabular}{cc}
		\subfloat[Over-damped model~\label{fig:power-overdamped}]{%
			\includegraphics[width=0.4\textwidth]{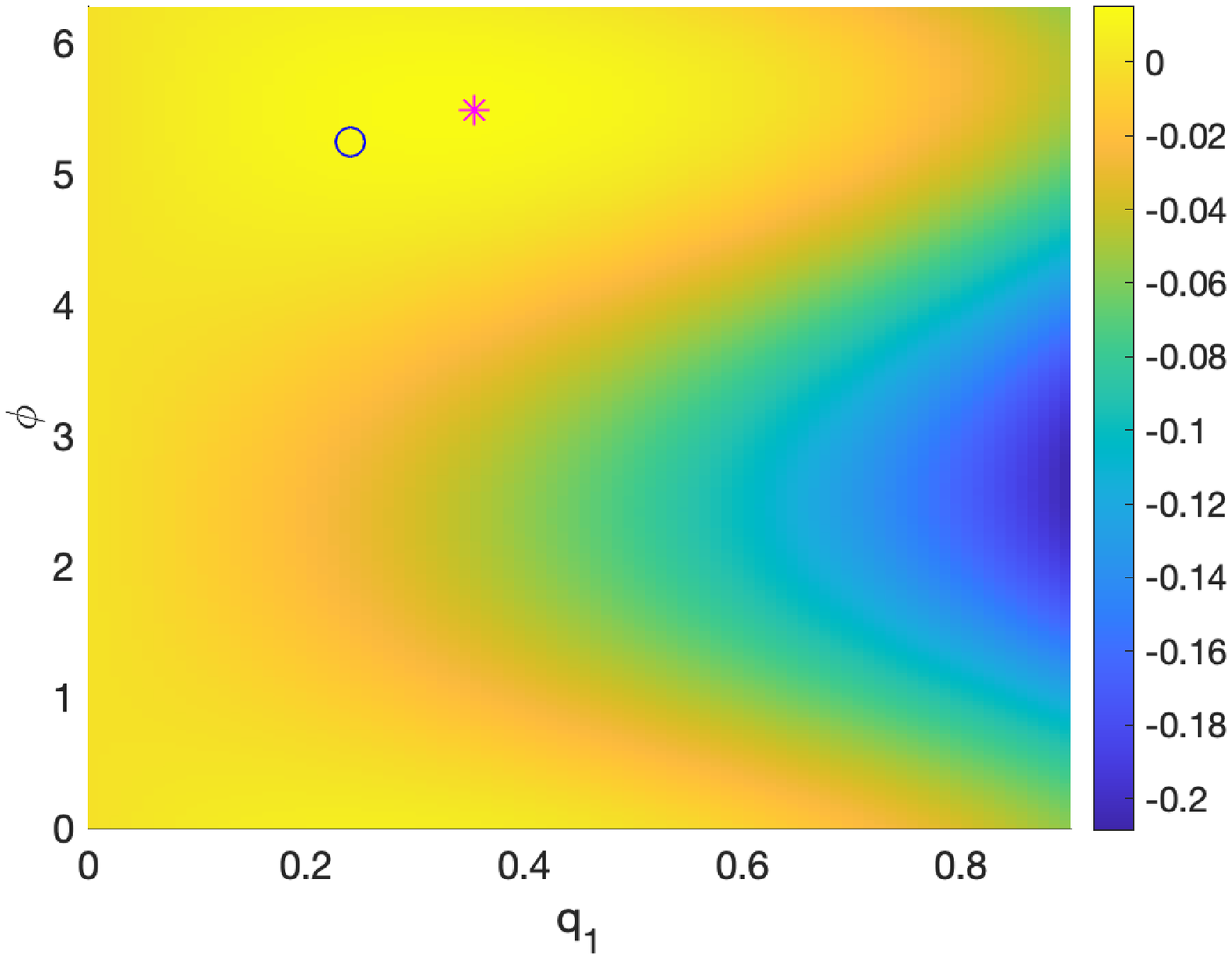}
		} &
		\subfloat[Under-damped model~\label{fig:power-underdamped}]{%
			\includegraphics[width=0.4\textwidth]{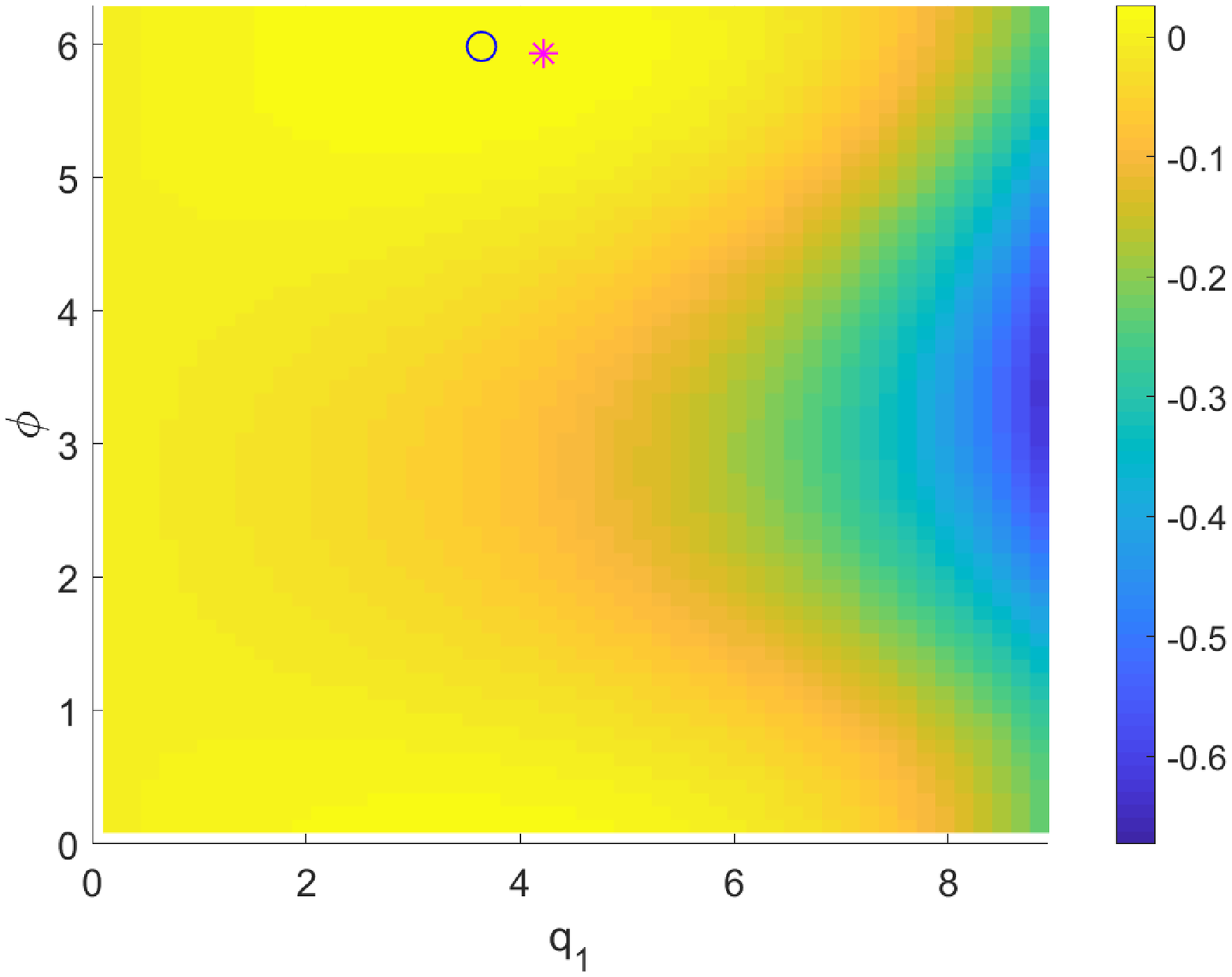}
		} 
	\end{tabular}
		\caption{Numerical evaluation of the power output as a function of the control phase $\phi$ and control amplitude $q_1$ as described in Section~\ref{sec:numerics}. 
			The points marked by "$\circ$" and  "$*$" correspond to optimal control parameters. The first ("$\circ$") was computed numerically, and the second ("$*$") analytically using \eqref{eq:q-phi-od} for the over-damped, and~\eqref{eq:q-phi-ud} for the under-damped model. } 
		\label{fig:power}
	\end{figure*}

	The numerical result for the power is depicted in Figure~\ref{fig:power-overdamped} and Figure~\ref{fig:power-underdamped} for the over-damped model and under-damped model respectively. It is observed that the pair of values $(\phi,q_1)$ that maximize the   power are close to the analytical expressions obtained by ignoring second-order terms. The model parameters used to obtain the numerical  result are presented in Table~\ref{tab:params}.
	
	\begin{table}[h]
		\centering
		\begin{tabular}{|c|c|c|c|}\hline
			Notation vs.\ value & notation &  overdamped& underdamped\\
			\hline 
			perturbation & $\epsilon$ &  1&1\\
			\hline
			viscosity coefficient
			&$\gamma$ & 1&1\\
			\hline
			frequency
			&$\omega$ & 2&2\\
			\hline
			temperature&$T_1$& 0.5& 0.5\\
			\hline
			temperature&$T_0$ & 1& 1\\
			\hline
			nominal gain
			&$q_0$& 1& 10$^*$\\
			\hline
		\end{tabular}
		\caption{Parameters selected in the simulations. Note that the value marked with {$^*$ is chosen to ensure stability.}}
		\label{tab:params}
	\end{table}
	
\begin{figure*}[t]
		\centering
		\begin{tabular}{cc}
		\subfloat[Over-damped model~\label{fig:eta-od}]{%
			\includegraphics[width=0.4\textwidth]{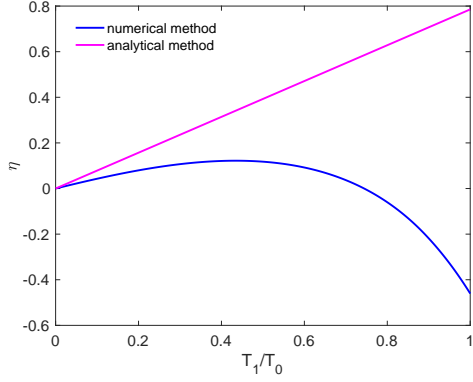}
		} &
		\subfloat[Under-damped model~\label{fig:eta-ud}]{%
			\includegraphics[width=0.4\textwidth]{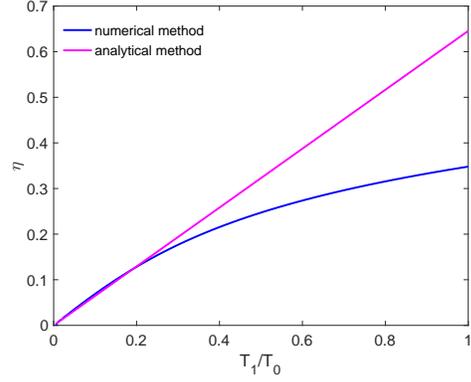}
		} 
	\end{tabular}
		\caption{Numerical evaluation of efficiency as a function of temperature  fluctuation $T_1$, at maximum power. The numerical result is compared with the analytical expressions derived using first-order approximations, in~\eqref{eq:eta-od} and~\eqref{eq:eta-ud} for the over-damped model and under-damped model, respectively.
		}
	\end{figure*}

	The thermal efficiency is also evaluated numerically, for the optimal values $\phi^*$  and $q^*$ that maximize the power. The efficiency as a function of the temperature fluctuation $T_1$ is depicted in Figures~\ref{fig:eta-od} and \ref{fig:eta-ud}, for the over-damped and under-damped cases, respectively. The numerical result is compared with the analytical result obtained up to first-order approximation. It is observed that the analytical expression captures the behaviour of the efficiency for small values of $T_1$ for both models.

	\section{Concluding remarks}\label{sec:CR}
	We addressed the question of maximal power and efficiency for thermodynamic processes in contact with a  heat bath having periodic continuously varying temperature. Our analysis is approximate and focuses on sinusoidal fluctuations. It is of interest to study the effect of the temperature profile, and properties of optimal controlling potential, in a more general setting.

\bibliographystyle{IEEEtran}
\bibliography{refs}

\end{document}